\documentclass{sf2a-conf2015}
\usepackage{graphicx}
\usepackage{hyperref}
\usepackage[]{natbib}  
\usepackage{epstopdf}
\usepackage{bm}
\usepackage{amsmath}
\usepackage{amssymb}

\def\BibTeX{{\rm B\kern-.05em{\sc i\kern-.025em b}\kern-.08em
    T\kern-.1667em\lower.7ex\hbox{E}\kern-.125emX}}
\bibpunct{(}{)}{;}{a}{}{,}  

\begin{document}

\TitreGlobal{SF2A 2015}


\title{Free inertial modes in differentially rotating convective envelopes of low-mass stars : numerical exploration}

\runningtitle{Free inertial modes in differentially rotating envelopes of low-mass stars}

\author{M. Guenel}\address{Laboratoire AIM Paris-Saclay, CEA/DSM/IRFU/SAp - Universit\'e Paris Diderot - CNRS, 91191 Gif-sur-Yvette, France}

\author{C. Baruteau}\address{IRAP, Observatoire Midi-Pyr\'en\'ees, Universit\'e de Toulouse, 14 avenue Edouard Belin, 31400 Toulouse, France}

\author{S. Mathis$^1$}
\author{M. Rieutord$^2$}



\setcounter{page}{237}


\maketitle


\begin{abstract}
Tidally-excited inertial waves in stellar convective regions are a key mechanism for tidal dissipation in stars and therefore the evolution of close-in binary or planetary systems. As a first step, we explore here the impact of latitudinal differential rotation on the properties of free inertial modes and identify the different families of modes. We show that they differ from the case of solid-body rotation. Using an analytical approach as well as numerical calculations, we conclude that critical layers --- where the Doppler-shifted frequency vanishes --- could play a very important role for tidal dissipation.
\end{abstract}



\section{Introduction}
Star-planet tidal interactions may result in the excitation of inertial waves in the convective region of stars. Their dissipation plays a prominent role in the long-term orbital evolution of short-period planets \citep{OgilvieLin2007, Lai2012, Valsecchi2014, Mathis2015}. Furthermore, turbulent convection sustains differential rotation in the envelope of low-mass stars, with an equatorial acceleration (as in the Sun) or deceleration which can modify the waves' propagation properties \citep{Brun2002, Brown2008,Gastine2014}. In this work, we explore the general properties of free linear inertial modes in a differentially rotating homogeneous isentropic fluid in a spherical shell. We assume that the angular velocity background flow depends on the latitudinal coordinate only, close to what is expected in the external convective envelope of low-mass stars. We use i) an analytical approach in the inviscid case to get the dispersion relation, study the existence of attractor cycles and identify the different families of inertial modes ; ii) high-resolution numerical calculations based on a spectral method for the viscous problem.

\section{Physical model}

We consider that the convective envelope of a low-mass star is a spherical shell of external radius $R$ and aspect ratio $\eta$ ($0<\eta<1$), extending from the boundary of the radiative core of radius $\eta R$ to the surface. For the sake of simplicity, we assume that the fluid is homogeneous with density $\rho_0$ and has a constant kinematic viscosity $\nu$.

\subsection{Hydrodynamic equations}

Since we want to study the propagation of inertial waves --- whose restoring force is the Coriolis acceleration --- in a differentially rotating fluid, we linearize the Navier-Stokes equations around the steady state where the fluid has a non-uniform dimensionless angular velocity $\Omega$. We normalize all frequencies by $\Omega_{\rm ref}$ which we define as the angular velocity at the poles and all distances by $R$. We thus look for dimensionless velocity perturbations ($\bf u$) and reduced pressure perturbations ($p$) of dimensionless angular frequency $\omega_p$ (in the inertial frame) and azimuthal wavenumber $m$. This means they are proportional to $\exp\left(i\omega_p t + im\varphi\right)$.

In an inertial frame, this yields the following system \citep[e.g.][]{BR2013} :
\begin{equation}
\begin{cases}
& i \tilde{\omega}_p {\bf u} + 2 \Omega {\bf e_z} \times {\bf u} + x \sin\theta \left( {\bf u}\cdot\nabla\Omega \right) {\bf e_{\varphi}} = -\nabla p + E \Delta {\bf u}, \\
& \nabla \cdot {\bf u} = 0,
\end{cases}
\label{eq:dimensionless_system}
\end{equation}
where $\tilde{\omega}_p = \omega_p + m\Omega $ is the dimensionless Doppler-shifted frequency, ${\bf e_z}$ is the unit vector along the rotation axis, $x$ is the dimensionless spherical radius and the Ekman number $E = \nu / R^2 \Omega_{\rm ref}$ is the dimensionless viscosity.

Note that we will numerically solve the vorticity equation
\begin{equation}
\begin{cases}
& \nabla \times \left( i \tilde{\omega}_p {\bf u} + 2 \Omega {\bf e_z} \times {\bf u} + x \sin\theta \left( {\bf u}\cdot\nabla\Omega \right) {\bf e_{\varphi}} \right) =  E \nabla \times \Delta {\bf u}, \\
& \nabla \cdot {\bf u} = 0,
\end{cases}
\label{eq:viscous_problem}
\end{equation}
in the following section in order to get rid of the $\nabla p$ term.

In addition to Eqs. (\ref{eq:viscous_problem}), we use stress-free boundary conditions (${\bf u} \cdot {\bf e_r} = 0$ and ${\bf e_r} \times [\sigma]{\bf e_r} = \bf 0$, where $[\sigma]$ is the viscous stress tensor) at the inner and outer boundaries of the spherical shell.

\subsection{Differential rotation profile}

As motivated in the introduction, we use a conical differential rotation profile that only depends on the colatitude $\theta$, which reads :
\begin{equation}
\Omega(\theta) = 1 + \varepsilon \sin^2\theta,
\label{eq:rotation_profile}
\end{equation}
so that the dimensionless angular velocity of the background flow is $1$ at the poles and $1+\varepsilon$ at the equator. The quantity $\varepsilon$ is a parameter that describes the behavior of the differential rotation :
\begin{itemize}
\item $\varepsilon > 0$ is for solar differential rotation (equatorial acceleration),
\item $\varepsilon < 0$ is for anti-solar differential rotation (equatorial deceleration).
\end{itemize}

\subsection{Dispersion relation and inviscid analysis}

When viscosity is neglected, Eqs. (\ref{eq:dimensionless_system}) is equivalent to a mixed-type second-order partial differential equation (PDE) for $p$ only. Then we can infer analytically the propagation properties of inertial waves in the inviscid limit through the solution's characteristic trajectories in the domains where this PDE is of hyperbolic type \citep{Rieutord2001, BR2013}. In solid-body rotation with angular velocity $\Omega_{\rm ref}$, the Doppler-shifted frequency is restricted to $[-2\Omega_{\rm ref},2\Omega_{\rm ref}]$ and these trajectories are straight lines with a fixed inclination angle $$\lambda=\sin^{-1}\left( \frac{\Omega_p + m\Omega_{\rm ref}}{2\Omega_{\rm ref}}\right)=\sin^{-1}(\tilde{\omega}_p/2)$$ with respect to the rotation axis. When differential rotation is included, we can numerically integrate these trajectories for any parameters using a ray tracing code. We found that they sometimes converge towards limit cycles called ``wave attractors'' that have previously been found in the case of solid-body rotation \citep{Rieutord2001}. Some of the figures shown in the following section feature overplotted white curves that were obtained using this method. We also found that the aforementioned PDE is not always hyperbolic in the whole shell, leading to different dynamics of the characteristic trajectories (latitudinal trapping, focusing towards a wedge, etc).

This analysis shows that two kinds of inertial modes may exist with differential rotation : D modes which can propagate in the entire spherical shell, and DT modes which can only propagate in part of the shell \citep{BR2013}. We also derive analytically the dispersion relation as well as the expressions of the phase and group velocities, and we show that when $m\neq0$, critical layers (or corotation resonances) defined by $\tilde{\omega}_p = 0$ are singularities of the inviscid problem. For more details, see Guenel et al. (2015, submitted to A\&A).

\begin{figure}
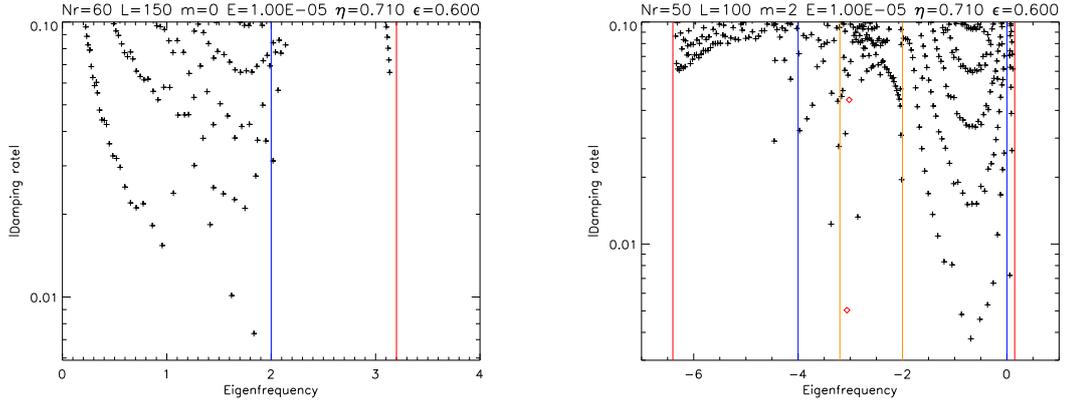

\centering
\includegraphics[width=0.4\textwidth]{{{guenel_fig1}}} \qquad
\includegraphics[width=0.4\textwidth]{{{guenel_fig2}}}
\caption{Distribution of eigenvalues in the complex plane for $m=0$ (left) and $m=2$ (right), with $E=10^{-5}$, $\eta=0.71$ and $\varepsilon=0.60$. Unstable eigenvalues with a positive damping rate are depicted by the red squares. The vertical blue (resp. red) line depicts the transition between the D and DT modes (resp. DT and non-existant modes) frequency ranges. Critical layers exist between the two orange lines.}
\label{fig:QZ}
\end{figure}

\section{Numerical calculations}
\subsection{Numerical method}

In this section, we show several results we obtained by solving numerically Eqs. (\ref{eq:viscous_problem}) --- along with stress-free boundary conditions --- using a unique decomposition of the unknown velocity field $\bf u$ onto vectorial spherical harmonics \citep{Rieutord1987} :
\begin{equation}
\label{eq:decomposition}
{\bf u}(x,\theta,\varphi) = \sum_{l=0}^{\infty} \sum_{m=-l}^{l} \left\{ u^l_m(x) {\bf R}^m_l(\theta,\varphi) + v^l_m(x) {\bf S}^m_l(\theta,\varphi) + w^l_m(x) {\bf T}^m_l(\theta,\varphi) \right\}
\end{equation}
with ${\bf R}^m_l = Y_l^m(\theta,\varphi){\bf e_r}$, ${\bf S}^m_l = {\bf \nabla_H} Y_l^m$ and ${\bf T}^m_l = {\bf \nabla_H} \times {\bf R}^m_l$ where $Y^m_l(\theta, \varphi)$ is the usual spherical harmonic of degree $l$ and order $m$ normalized on the unit sphere, and ${\bf \nabla_H} = {\bf e_{\theta}} \, \partial_{\theta} + {\bf e_{\varphi}} (\sin\theta)^{-1} \, \partial_{\varphi} $ is the horizontal gradient.

This decomposition yields a linear system of coupled ordinary differential equations involving $u^l_m$, $v^l_m$, $w^l_m$, which can be discretized in the radial direction on the Gauss-Lobatto collocation nodes ($N_r$ points). We also truncate the system at a maximum spherical harmonic degree $L$. The linear solver we use \citep[see][]{Rieutord1987} can compute directly the set of eigenvalues of the sparse matrix for moderate resolutions. It can also be used to compute individual pairs of eigenvalues and eigenmodes at higher resolutions.

\subsection{Exploration of the modes}

\begin{figure}
\centering
\includegraphics[width=0.35\textwidth]{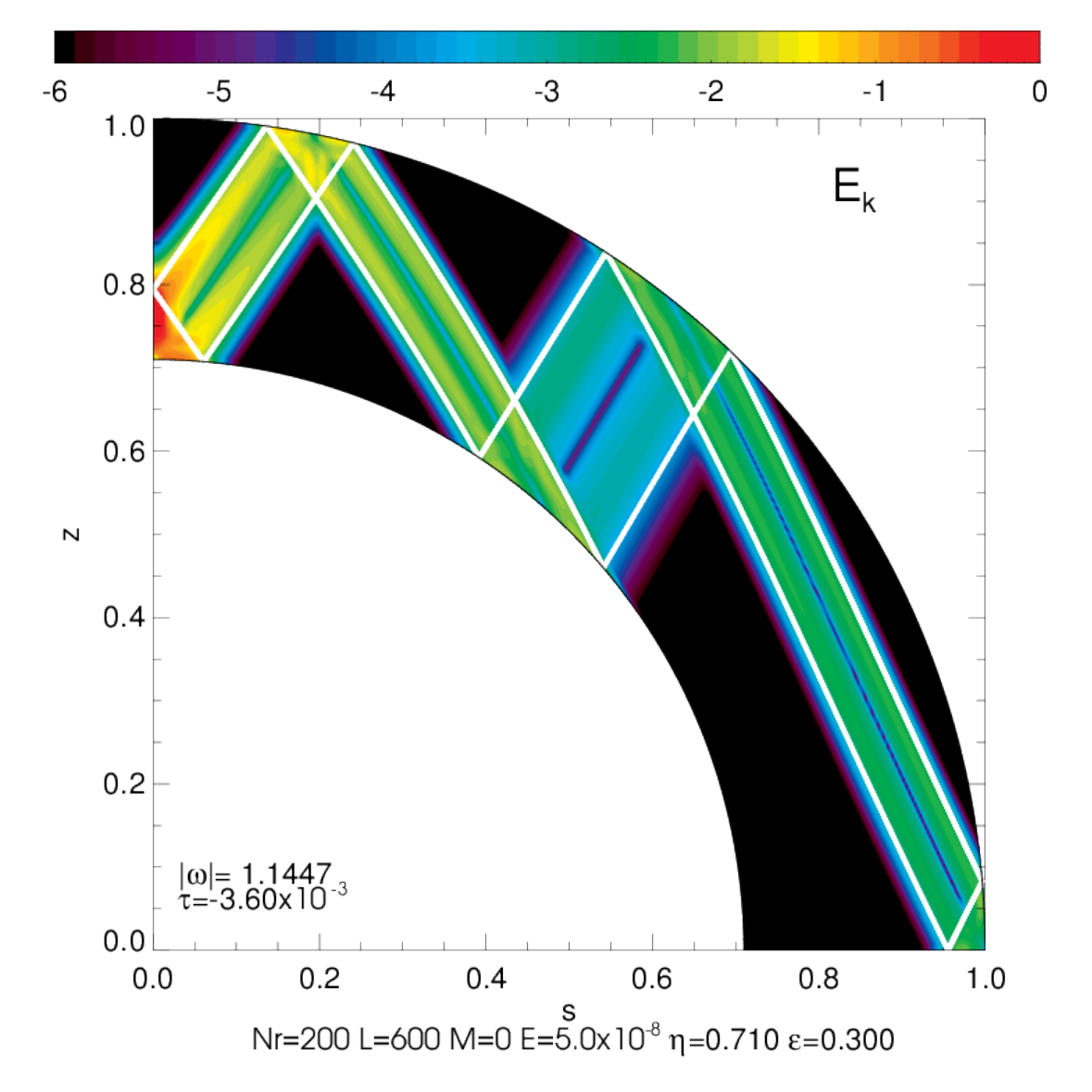} \qquad
\includegraphics[width=0.35\textwidth]{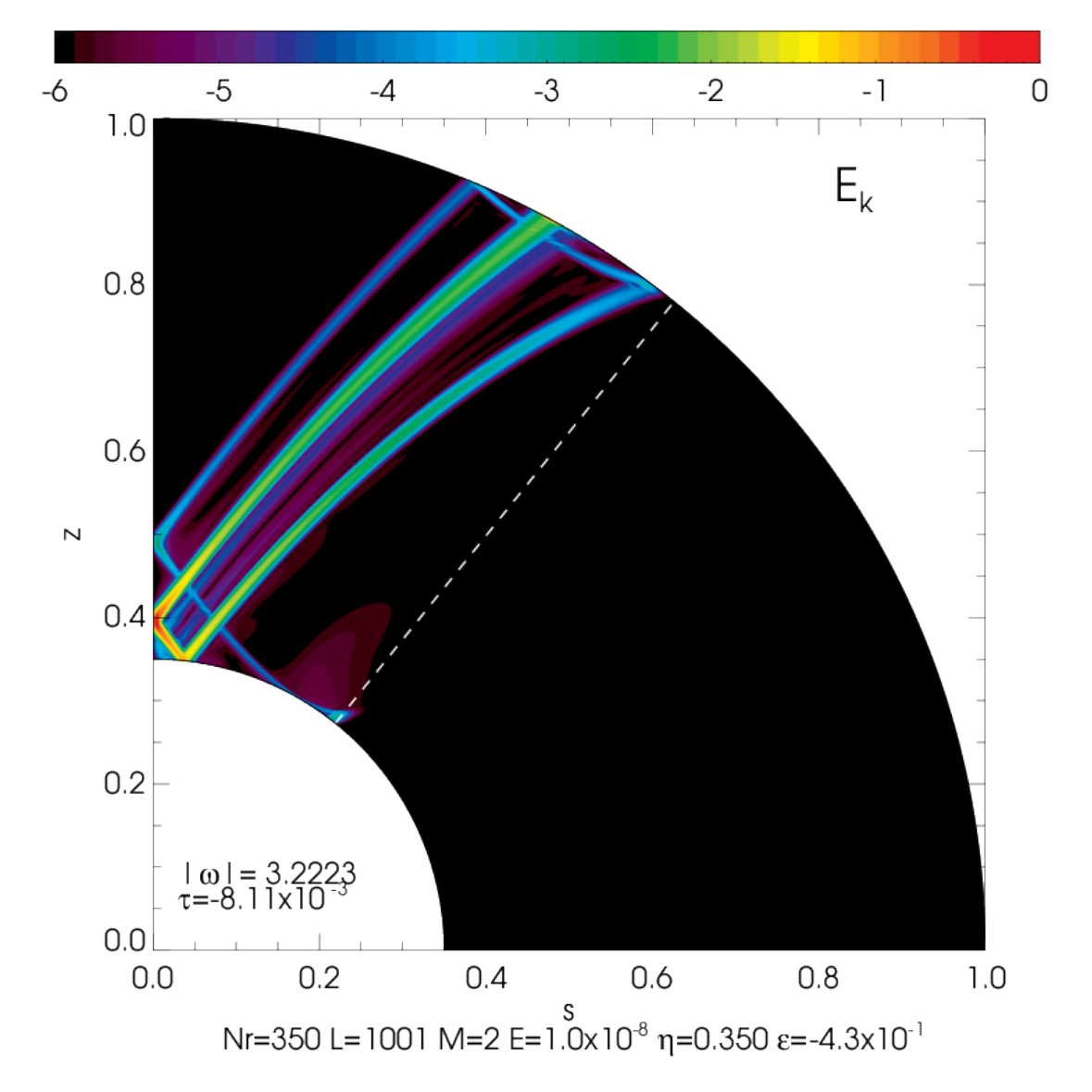}
\includegraphics[width=0.35\textwidth]{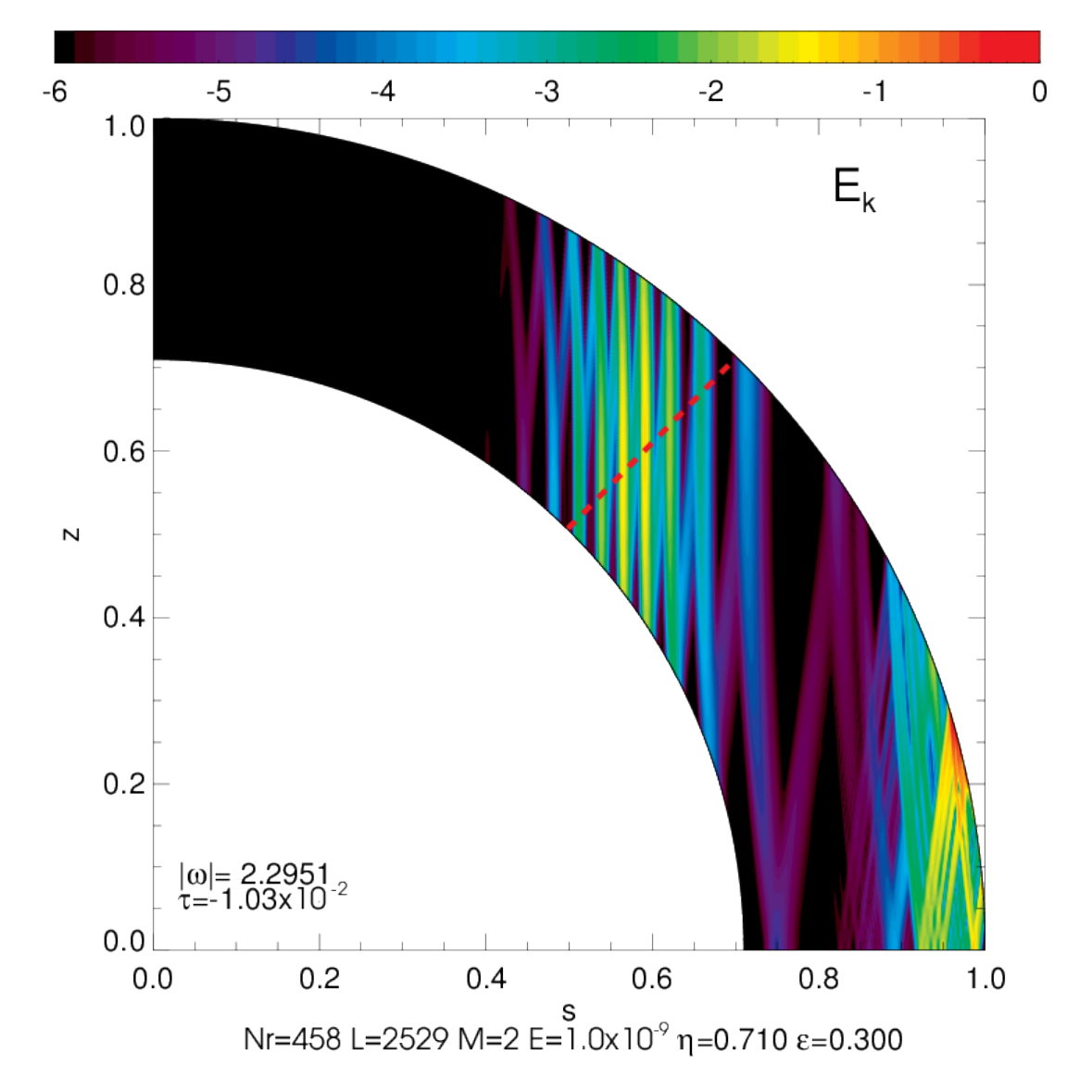} \qquad
\includegraphics[width=0.35\textwidth]{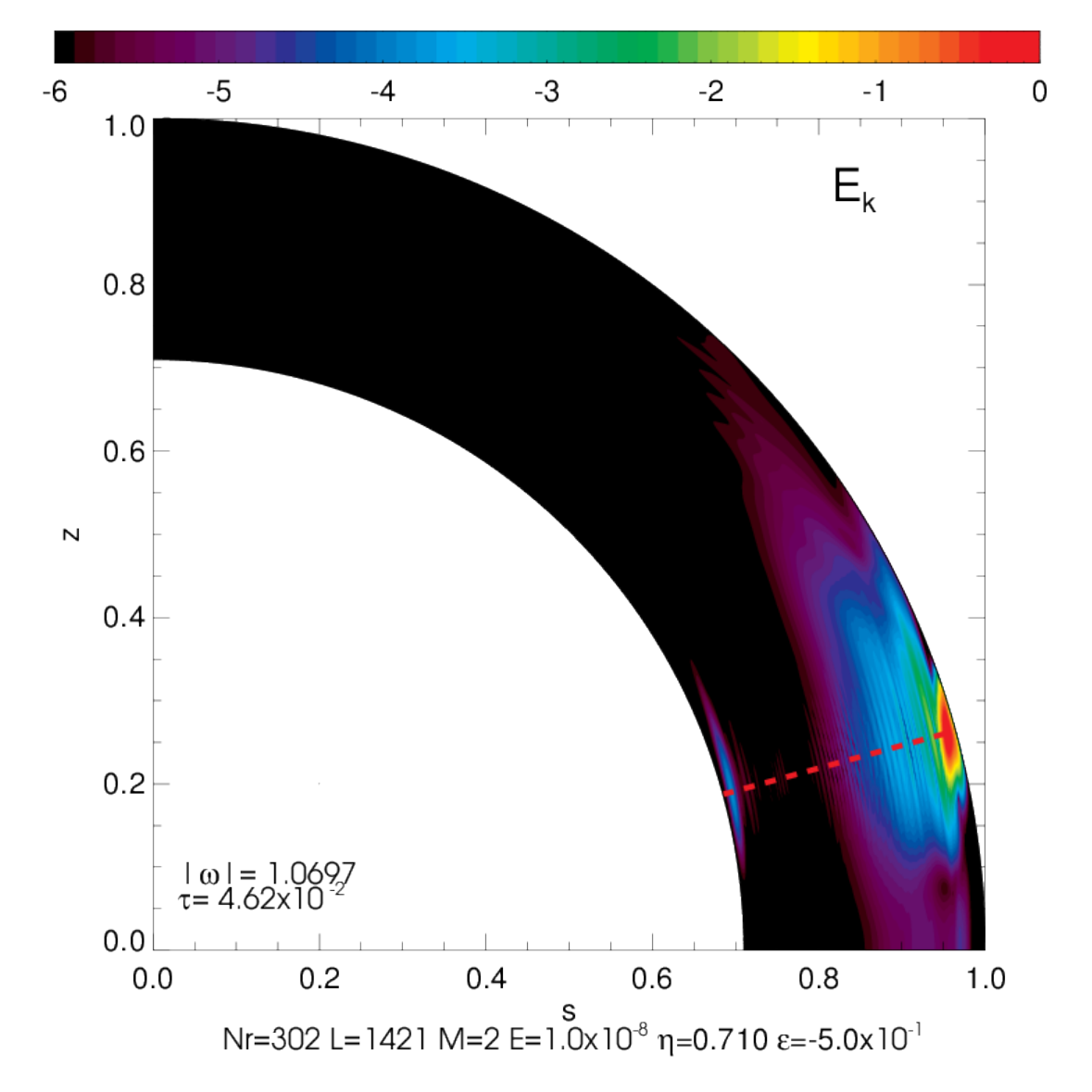}
\caption{Meridional cuts of the normalized kinetic energy of different inertial modes. {\bf Top-left~:} Axisymmetric D mode. The attractor of characteristics for these parameters is overplotted by the white curve. {\bf Top-right~:} DT mode with focusing towards a wedge at the intersection of a turning surface, depicted by the white-dashed line, with the inner boundary of the shell. {\bf Bottom-left~:} Stable non-axisymmetric D mode with critical layer (depicted by the red-dashed line). {\bf Bottom-right~:} Unstable non-axisymmetric D mode with corotation resonance.}
\label{fig:modes}
\end{figure}
As a first step, we computed numerically the entire set of eigenvalues using a QZ-factorization method at moderate resolutions and therefore Ekman numbers $E$ above $10^{-6}$. The algorithm is designed to obtain the least-damped modes, which means that eigenvalues whose absolute damping rate exceeds $\sim 10^{-1}$ can be safely ignored. In Fig. \ref{fig:QZ}, we show the results of two of these computations for a star with the solar aspect ratio $\eta=0.71$ and $\varepsilon=0.60$ (the Sun is actually around $\varepsilon=0.30$), for $m=0$ (left) and $m=2$ (right). The main results we obtained from these computations are the following :
\begin{itemize}
\item the least-damped modes are always in the frequency range that corresponds to D modes ;
\item eigenvalues in the DT frequency range are rare and have higher absolute damping rates ;
\item when $m\neq0$ and a critical layer exists in the shell, unstable eigenvalues exist for $E$ as high as $10^{-5}$.
\end{itemize}

Then, we computed various eigenmodes for different parameters , as shown by the few examples in Fig. \ref{fig:modes}. We briefly sum up our findings below :
\begin{itemize}
\item D modes have properties that are very similar to inertial modes in the case of solid-body rotation : characteristic trajectories are curved but wave attractors still exist in narrow frequency bands and patterns of shear layers form around them (see the top-left panel of Fig. \ref{fig:modes}). Their damping rate usually scales as $E^{1/3}$ as in solid-body rotation.
\item DT modes with small damping rates are rare as indicated by the diagrams shown in the previous paragraph. Moreover, the shear layers that compose them often focus towards the intersection of a turning surface with the inner or outer boundary of the shell (see the top-right panel of Fig. \ref{fig:modes}).
\item When a corotation resonance exists inside the shell, both stable and unstable modes exist (see the bottom panels of Fig. \ref{fig:modes}). The characteristic trajectories and the shear layers that follow them may become vertical at the resonance with a local accumulation of kinetic energy. On the other hand, unstable modes show no recognizable shear layer patterns and they may play a prominent role in the dissipation and/or exchange of angular momentum with the mean flow \citep[see][]{Grimshaw1979,Watson1981}.
\end{itemize}

\section{Conclusions}

We find that modes that can propagate in the whole differentially rotating convective envelope of a low-mass star behave the same way as with solid-body rotation. However, other families of inertial modes exist, which can propagate only in a restricted part of the convective zone. Most importantly, non-axisymmetric oscillation modes may be unstable when a critical layer exists in the convective zone. These new properties of free inertial modes with differential rotation could significantly change our understanding of the tidal dissipation and the dynamics of short-period systems. This is related to the tidally-forced regime which we will study in the near future.

\begin{acknowledgements}
S. Mathis and M. Guenel acknowledge funding by the European Research Council through ERC grant SPIRE 647383. This work was also supported by the Programme National de Planétologie (CNRS/INSU) and the CoRoT-CNES grant at Service d?Astrophysique (CEA Saclay).
\end{acknowledgements}

\bibliographystyle{aa}  
\bibliography{guenel} 

\end{document}